\documentstyle[epsf,multicol,aps]{revtex}

\newcommand{\vecS}{\mbox{\boldmath $S$}}
\newcommand{\vecM}{\mbox{\boldmath $M$}}
\newcommand{\vecr}{\mbox{\boldmath $r$}}
\newcommand{\vece}{\mbox{\boldmath $e$}}
\newcommand{\e}{\epsilon}

\begin{document}
\draft
\title{Probability-Changing Cluster Algorithm for Two-Dimensional 
XY and Clock Models}

\author{Yusuke Tomita\cite{tomita} and Yutaka Okabe\cite{okabe}}

\address{
Department of Physics, Tokyo Metropolitan University,
Hachioji, Tokyo 192-0397, Japan
}

\date{Received \today}

\maketitle

\begin{abstract}
We extend the newly proposed probability-changing cluster 
(PCC) Monte Carlo algorithm to the study of systems with 
the vector order parameter.  
Wolff's idea of the embedded cluster formalism is used 
for assigning clusters.  
The Kosterlitz-Thouless (KT) transitions for the two-dimensional (2D) 
XY and $q$-state clock models are studied by using the PCC algorithm.  
Combined with the finite-size scaling analysis based on the KT form of 
the correlation length, $\xi \propto \exp(c/\sqrt{T/T_{\rm KT}-1})$, 
we determine the KT transition temperature and the decay exponent $\eta$ as 
$T_{\rm KT}=0.8933(6)$ and $\eta=0.243(5)$ for the 2D XY model. 
We investigate two transitions of the KT type 
for the 2D $q$-state clock models with $q=6,8,12$, 
and {\it for the first time} 
confirm the prediction of $\eta = 4/q^2$ at $T_1$, 
the low-temperature critical point between the ordered and XY-like 
phases, systematically. 
\end{abstract}


\begin{multicols}{2}
\narrowtext

\section{Introduction}

The two-dimensional (2D) XY model shows a unique phase transition, 
the Kosterlitz-Thouless (KT) transition \cite{KT,Kosterlitz}.  
It does not have a true long-range order, but the correlation function 
decays as a power of the distance at all the temperatures 
below the KT transition point. 
Jos{\'e} {\it et al} \cite{Jose} studied the effect of the 
$q$-fold symmetry-breaking fields on the 2D XY model;  this is 
essentially the same as treating the $q$-state clock model, 
where only the discrete values are allowed for the angle $\theta_i$ 
of the XY spins such that 
\begin{equation}
 \theta_i = 2\pi p_i/q \quad {\rm with} \quad p_i=0, 1, 2, \cdots, q-1.
\label{theta}
\end{equation}
In the limit $q \rightarrow \infty$ 
we get the XY model.
It was shown that the 2D $q$-state clock model has 
two phase transitions of the KT type at $T_1$ and $T_2$ 
($T_1<T_2$) for $q>4$.  There is an intermediate XY-like phase 
between a low-temperature ordered phase ($T<T_1$) 
and a high-temperature disordered phase ($T>T_2$).  
It was predicted that the decay critical exponent $\eta$ varies from
$\eta=1/4$ at $T_2$ to $\eta=4/q^2$ at $T_1$.

Most of the above theoretical analyses relied on the renormalization 
group argument, and they are not exact.  
There have been extensive numerical studies on the 2D classical 
XY model \cite{Weber,Gupta,Kawamura,Butera,Olsson,Zheng,Chung}. 
In contrast, only a limited number of numerical works have been 
reported on the $q$-state clock model 
\cite{Challa,Yamagata}, 
and the accuracy was not good enough 
especially for the low-temperature phase transition. 
There have been no systematic studies to check the prediction 
of $\eta(T_1) = 4/q^2$.

In numerical studies, efficient algorithms are important 
for getting the necessary information.  
The cluster update algorithms of the Monte Carlo simulation 
\cite{SwWa,Wolff89} are examples of such efforts, and 
they are useful for overcoming the problems of slow dynamics. 
Recently we proposed an effective cluster algorithm,
which is called the probability-changing cluster (PCC) algorithm,
of tuning the critical point automatically \cite{PCC}.  
It is an extension of the Swendsen-Wang algorithm \cite{SwWa}, 
but we change the probability of cluster update 
(essentially, the temperature) depending on 
the observation whether clusters are percolating or not percolating. 
We showed the effectiveness of the PCC algorithm for the Potts 
models \cite{PCC}, determining the critical point and 
critical exponents for the second-order phase transition 
precisely with less numerical efforts. 
The PCC algorithm was also applied to the 2D site-diluted 
Ising model, where the crossover and the self-averaging 
properties were studied \cite{to01a}. 
The advantage of using the PCC algorithm in the study of random systems 
is as follows:  The sample-dependent $T_c(L)$ for the finite system 
with the linear size $L$ is important for taking sample average, 
and the PCC algorithm is suitable for getting 
the sample-dependent $T_c(L)$. 

There are a lot of interesting questions about the extension of 
the PCC algorithm.  (i) Can the PCC algorithm be used for 
the problem of the vector order parameter, such as the XY model? 
(ii) Can it be applied to the analysis of 
the transition of the KT type?  
(iii) Can it work even if the system shows two or more phase transitions?

The purpose of the present paper is to answer these questions. 
We extend the PCC algorithm so as to treat 
systems with the vector order parameter. 
The rest of the paper is organized as follows. 
In Sec.~II, we formulate the extension of the PCC algorithm 
for the vector order parameter.  In Sec.~III, the KT transition of 
the 2D XY model is studied with the finite-size scaling (FSS) 
analysis based on the KT form of the correlation length. 
In Sec.~IV, we study the KT transitions of the 2D clock models. 
We investigate both phase transitions at $T_1$ and $T_2$ for 
the $q=6,8,12$ clock models. 
The summary and discussions are given in Sec.~V. 

\section{PCC algorithm for vector order parameter}

Our Hamiltonian is given by 
\begin{equation}
 {\cal H} = -J\sum_{\left<i,j\right>} \vecS_i \cdot \vecS_j, 
\label{H_XY}
\end{equation}
where $\vecS_i$ is a planar unit vector, 
$(\cos \theta_i, \sin \theta_i)$, at site $i$; 
$\theta_i$ takes the value of $[0,2\pi)$ for the XY model, 
and the value given in Eq.~(\ref{theta}) for the $q$-state clock model. 
The summation is taken over the nearest-neighbor pairs $\left<i,j\right>$. 

In order to extend the PCC algorithm to systems 
with the vector order parameter, 
we use Wolff's idea of the embedded cluster formalism \cite{Wolff89}. 
We project the vector $\vecS_i$ onto a randomly chosen 
unit vector $\vece_1$ and another unit vector $\vece_2$, 
perpendicular to $\vece_1$, as
\begin{equation}
 \vecS_i = \vece_1 \cos \phi_i + \vece_2 \sin \phi_i, 
 \label{project}
\end{equation}
where $\phi_i$ is the angle measured from the axis of 
the vector $\vece_1$. 
Then, the Hamiltonian, Eq.~(\ref{H_XY}), is rewritten as
\begin{equation}
 {\cal H} = - \sum_{\left<i,j\right>} 
  (J_{ij}^{(1)}  \e_i^{(1)} \e_j^{(1)} +
   J_{ij}^{(2)}  \e_i^{(2)} \e_j^{(2)} )
\end{equation}
with positive effective couplings
\begin{equation}
 J_{ij}^{(1)} = J \, |\cos \phi_i| \, |\cos \phi_j|, \quad
 J_{ij}^{(2)} = J \, |\sin \phi_i| \, |\sin \phi_j| 
\end{equation}
for two sets of Ising variables $\{ \e_i^{(1)} \}$ and $\{ \e_i^{(2)} \}$.
Formally, we can restrict ourselves to the region $[0, \pi/2)$ for 
$\{ \phi_i \}$, and we write the partition function as 
\begin{eqnarray}
 Z = \int_0^{\pi/2} \{ d\phi_i \} 
     & & \hspace{-4mm} \sum_{\{ \e_i^{(1)}=\pm 1 \}} 
     \exp( \beta \sum_{\left<i,j\right>} J_{ij}^{(1)} \, 
     \e_i^{(1)} \e_j^{(1)}) \nonumber \\
     & \times &  \sum_{\{ \e_i^{(2)}=\pm 1 \}}
     \exp( \beta \sum_{\left<i,j\right>} J_{ij}^{(2)} \, 
     \e_i^{(2)} \e_j^{(2)}) 
\end{eqnarray}
with $\beta = 1/k_BT$.  Then, we can use 
the Kasteleyn-Fortuin (KF) cluster representation 
for the Ising spins \cite{KF}. 
To make the KF cluster, we connect the bonds of parallel Ising spins 
with the probability
\begin{equation}
  p_{ij}^{(1,2)} = 1 - \exp(-2 \beta J_{ij}^{(1,2)}). 
\label{KF_p}
\end{equation}

In the PCC algorithm \cite{PCC}, the cluster representation 
of the Ising model is used in two ways. 
First, we flip all the spins on any KF cluster to one of two states, 
that is, $+1$ or $-1$, as in the Swendsen-Wang algorithm \cite{SwWa}.
Second, we change the KF probability, Eq.~(\ref{KF_p}), 
depending on the observation whether clusters are percolating or not. 
It is based on the fact that the spin-spin correlation function 
$G(\vecr_i-\vecr_j)$ becomes nonzero for $|\vecr_i-\vecr_j| 
\to \infty$ at the same point as the percolation threshold.  
For the XY model in the embedded cluster formalism, 
the spin-spin correlation function is written as
\begin{eqnarray}
  G(\vecr_i-\vecr_j) &=& \left< \vecS_i \cdot \vecS_j \right> 
 = \left< |\cos \phi_i| \, |\cos \phi_j| \, \e_i^{(1)} \e_j^{(1)} \right> 
 \nonumber \\
 &~& + \left< |\sin \phi_i| \, |\sin \phi_j| \, \e_i^{(2)} \e_j^{(2)} \right> 
 \nonumber \\ 
 &=& \left< A_{ij} \, \Theta^{(1)}(\vecr_i, \vecr_j) \right>
   + \left< B_{ij} \, \Theta^{(2)}(\vecr_i, \vecr_j) \right>,
\end{eqnarray}
where 
$\left< \cdots \right>$ represent the thermal average. 
The function $\Theta(\vecr_i, \vecr_j)$ is equal to 1 (0)
if the sites $i$ and $j$ belong to the same (different) cluster, 
and $A_{ij}$ and $B_{ij}$ are some constants. 
Thus, the system is regarded as percolating if $\e^{(1)}$ 
{\it or} $\e^{(2)}$ Ising spins are percolating. 
When treating the cluster spin update, one may consider 
Ising spins of a single type projected onto a randomly chosen axis, 
as in the original Wolff's proposal \cite{Wolff89}.  However, 
we should consider Ising spins of both types perpendicular to each other 
for checking the percolation.

The procedure of Monte Carlo spin update is as follows:
(i) Start from some spin configuration and some value of $\beta$. 
(ii) Choose a unit vector $\vece_1$ randomly. 
(iii) Construct the KF clusters for $\e^{(1)}$ and $\e^{(2)}$ 
using the probability, Eq.~(\ref{KF_p}), and check 
whether the system is percolating or not. 
Flip all the spins on any KF cluster to $+1$ or $-1$ for 
both $\e^{(1)}$ and $\e^{(2)}$ Ising spins.  
(iv) If the system is percolating (not percolating), 
decrease (increase) $\beta$ by $\Delta \beta \ (>0)$.
(v)
Go back to the process ii.

As $\Delta \beta$ becomes small, 
the distribution of $\beta$ becomes a sharp Gaussian 
distribution around the mean value $\beta_c(L)$, 
which depends on the system size $L$. 
We approach the canonical ensemble in this limit, 
and the existence probability $E_p$, the probability that 
the system percolates, becomes 1/2 at $\beta_c(L)$.  

\section{XY model}

We have made simulations for the classical XY model 
on the square lattice with the system sizes 
$L$ =8, 16, 32, 64, 128, 256, and 512.  
After 20,000 Monte Carlo sweeps of determining $\beta_c(L)$ 
with gradually reducing $\Delta \beta$, 
we have made 10,000 Monte Carlo sweeps to take the thermal average; 
we have made 100 runs for each size to get better statistics and 
to evaluate the statistical errors. 
As for the criterion to determine percolating,   
we have employed the topological rule \cite{PCC,machta} in the present study. 
The topological rule is that some cluster winds around the system 
in at least one of the $D$ directions in $D$-dimensional systems. 

\begin{figure}
\epsfxsize=0.95\linewidth
\centerline{\epsfbox{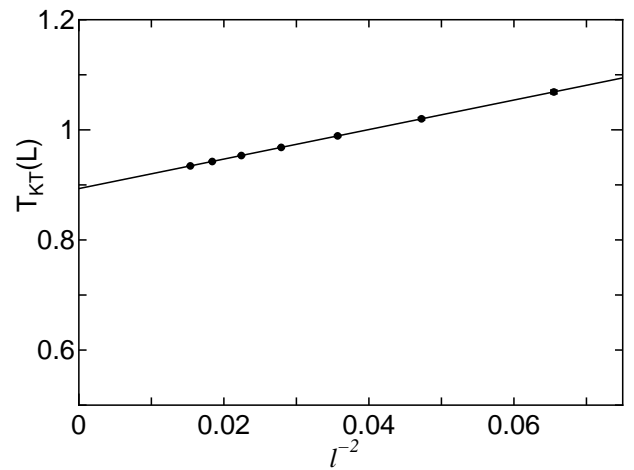}}
\caption{Plot of $T_{\rm KT}(L)$ 
of the 2D XY model for $L$ = 8, 16, 32, 64, 128, 256, and 512, 
where $l = \ln bL$. 
} 
\label{fig_1}
\end{figure}

Let us start with the size dependence of the transition temperature. 
We use the FSS analysis based on the KT form of 
the correlation length, 
\begin{equation}
 \xi \propto \exp(c/\sqrt{t}) 
\end{equation}
with $t=(T-T_{\rm KT})/T_{\rm KT}$. 
Using the PCC algorithm, 
we locate the temperature $T_{\rm KT}(L) = 1/k_B \beta_c(L)$
that the existence probability $E_p$ is 1/2.  
Then, using the FSS form of $E_p$, that is, $E_p=E_p(\xi/L)$, 
we have the relation 
\begin{equation}
 T_{\rm KT}(L) = T_{\rm KT} + \frac{c^2 T_{\rm KT}}{(\ln bL)^2}. 
\label{T_KT}
\end{equation}
We plot $T_{\rm KT}(L)$ as a function of $l^{-2}$ with 
$l = \ln bL$ for the best fitted parameters in Fig.~\ref{fig_1}.  
We represent the temperature in units of $J/k_B$.
The error bars are smaller than the size of marks. 
Our estimate of $T_{\rm KT}$ is 0.8933(6); the number 
in the parentheses denotes the uncertainty in the last digits. 
We have estimated the uncertainty by the $\chi^2$ test of the data 
for 100 samples. 
This value is consistent with the estimates of recent studies; 
0.89213(10) by the Monte Carlo simulation \cite{Olsson}, 
and 0.894 by the short-time dynamics \cite{Zheng}. 
The constant $c$, in Eq.~(\ref{T_KT}), is estimated as $c$=1.73(2). 

\begin{figure}
\epsfxsize=0.95\linewidth
\centerline{\epsfbox{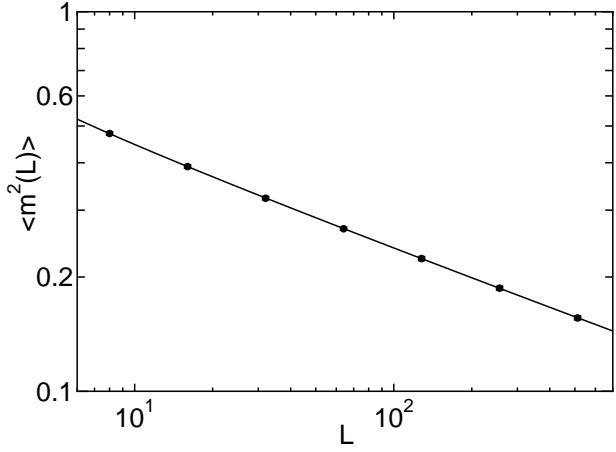}}
\caption{Logarithmic plot of $\left< m^2(L) \right>$ 
at $T_{\rm KT}(L)$ of the 2D XY model 
for $L$ = 8, 16, 32, 64, 128, 256, and 512. 
} 
\label{fig_2}
\end{figure}

Let us consider the magnetization 
$\left< m^2(L) \right>$ at $T_{\rm KT}(L)$ 
to discuss the critical exponent $\eta$. 
In Fig.~\ref{fig_2}, we plot $\left< m^2(L) \right>$ 
as a function of $L$ in logarithmic scale. 
We expect the FSS of the form $\left< m^2(L) \right> 
\propto L^{-\eta}$, but there are small corrections. 
The importance of the multiplicative logarithmic 
corrections were pointed out \cite{Kosterlitz,Janke97}. 
Using the form
\begin{equation}
 \left< m^2(L) \right> = A L^{-\eta} \, (\ln b'L)^{-2r}, 
\label{eta_1}
\end{equation}
we obtain $\eta= 0.243(5)$ and $r=0.038(5)$. 
We show the fitting curve obtained by using Eq.~(\ref{eta_1}) 
in Fig.~\ref{fig_2}.  This value of $\eta$ is a little bit 
smaller than the theoretical prediction, 1/4 (=0.25).  
Our logarithmic-correction exponent $r$ 
is compatible with Janke's result $r=0.0560(17)$ 
for thermodynamic data \cite{Janke97}, 
but different from the theoretical prediction 
$r=-1/16$ \cite{Kosterlitz}. 

\section{Clock model}

\begin{figure}
\epsfxsize=0.95\linewidth
\centerline{\epsfbox{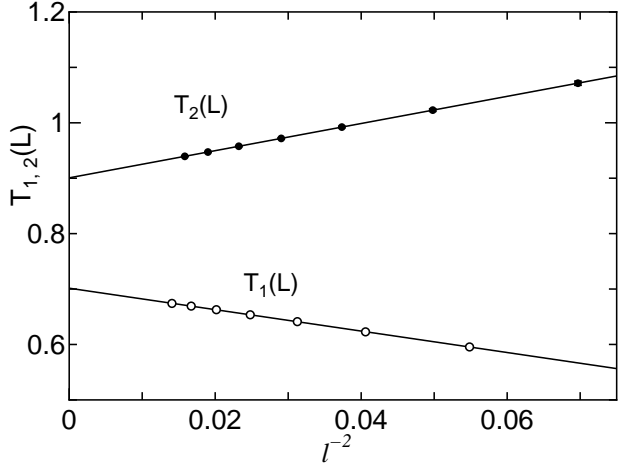}}
\caption{Plots of $T_{\rm 1}(L)$ and $T_{\rm 2}(L)$ 
of the 2D 6-state clock model for $L$ = 8, 16, 32, 64, 128, 256, and 512, 
where $l = \ln bL$. 
} 
\label{fig_3}
\end{figure}

\begin{figure}
\epsfxsize=0.95\linewidth
\centerline{\epsfbox{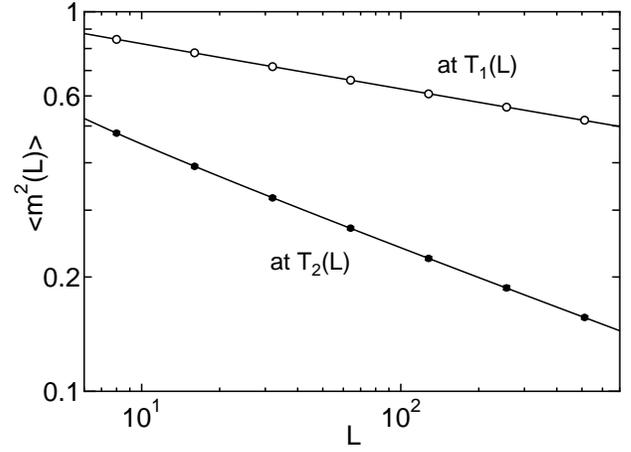}}
\caption{Logarithmic plots of $\left< m^2(L) \right>$ 
at $T_1(L)$ and $T_2(L)$ of the 2D 6-state clock model 
for $L$ = 8, 16, 32, 64, 128, 256, and 512. 
} 
\label{fig_4}
\end{figure}

Next turn to the $q$-state clock model. 
Because of the reflection symmetry, we confine ourselves to 
the case of even $q$.  Then, the same procedure can be 
used as the XY model.  One thing we should have in mind is that 
the axis of the vector $\vece_1$ should be chosen from 
one of $q$ directions in Eq.~(\ref{theta}) or the middle of 
two of them. 
We plot the high-temperature transition temperature $T_2(L)$ 
of the 6-state clock model as a function of $l^{-2}$ 
in Fig.~\ref{fig_3}.  
The estimate of $T_2$ is 0.9008(6), which is more precise than 
the previous estimates; 0.92(1) \cite{Challa} and 0.90 \cite{Yamagata}. 
The plot of $\left< m^2(L) \right>$ at $T_2(L)$ as a function of $L$ 
for the 6-state clock model is given in Fig.~\ref{fig_4}. 
The estimate of $\eta$ is 0.243(5) by the analysis of 
the multiplicative logarithmic corrections, Eq.~(\ref{eta_1}), 
and the exponent $r$ is estimated as 0.037(5). 

For the second-order transition, the curves of the existence probability 
$E_p$ of different sizes cross at $T_c$ as far as the corrections 
to FSS are negligible; this is the same as the behavior of 
the Binder ratio \cite{Binder}.  
For the KT transition, however, $T_{\rm KT}$ is not the crossing point 
but the spray-out point.  Therefore, $T_2$ can be searched only from 
the high-temperature side, and $T_1$ only from 
the low-temperature side.  The value of $E_p$ at $T_1$ is 
close to 1.  In principle, we can use the same 
procedure as the study of $T_2$; we may change the setting 
value of $E_p$, 1/2, to a higher one by introducing 
a biased random walk.  However, it is difficult to resolve 
the size dependence for lower temperatures. 
Therefore, we employ a slightly different approach 
for the analysis of the phase transition at $T_1$.  
When judging whether clusters are percolating or not, 
we consider another type of clusters. 
Instead of choosing the vector $\vece_1$ randomly 
in Eq.~(\ref{project}), we choose the vector $\vece_1$ as 
\begin{equation}
 \vecM = |\vecM| \, \vece_1 
\end{equation}
with $\vecM = \sum_i \vecS_i$,
or more generally we may choose $\vece_1$ such that
\begin{equation}
  \vecM = |\vecM| \, (\vece_1 \cos \phi + \vece_2 \sin \phi)
 \label{phi}
\end{equation}
with some fixed angle $\phi$. 
With this choice, the existence probability for the percolation of 
only $\e^{(1)}$ (or $\e^{(2)}$) Ising spins holds the 
same FSS property as the total $E_p$.  
As a result, we can control the value of $E_p^{(1)}$ at $T_1$ 
so as to apply the FSS analysis easily 
with an appropriate $\phi$.  

The low-temperature transition temperature $T_1(L)$ 
of the 6-state clock model obtained by the above 
modified approach is plotted as a function of $l^{-2}$ 
also in Fig.~\ref{fig_3}.  As the angle $\phi$ in Eq.~(\ref{phi}), 
we have used $\pi/3$. 
Our estimate of $T_1$ is 0.7014(11), which is more precise again than 
the previous estimates; 0.68(2) \cite{Challa} and 0.75 \cite{Yamagata}.
The plot of $\left< m^2(L) \right>$ at $T_1(L)$ 
for the 6-state clock model is also given in Fig.~\ref{fig_4}.
The estimate of $\eta$ is 0.113(3) by the analysis of 
the multiplicative logarithmic corrections, Eq.~(\ref{eta_1}), 
and the exponent $r$ is estimated as 0.017(4). 
The previous estimates of $\eta$ are 0.100 \cite{Challa} 
and 0.15 \cite{Yamagata}.

\begin{figure}
\epsfxsize=0.95\linewidth
\centerline{\epsfbox{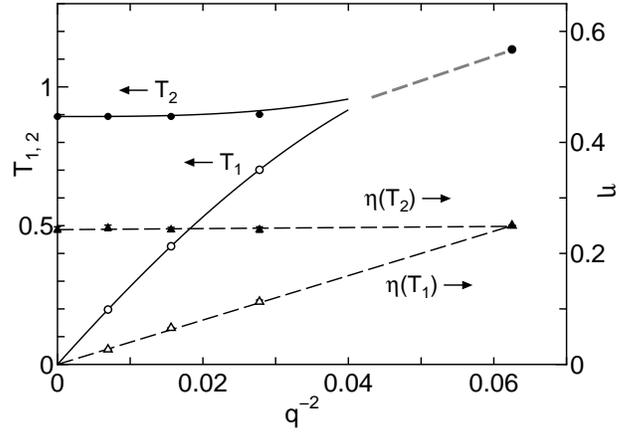}}
\caption{Transition temperatures and critical exponents 
as a function of $1/q^2$ for the 2D $q$-state clock model. 
} 
\label{fig_5}
\end{figure}

We have also made simulations for $q=8$ and $q=12$. 
The estimates of the transition temperatures $T_1, T_2$ 
and those of $\eta (T_1)$ and $\eta (T_2)$ for $q$=6, 8, 12, and 
$\infty$ (the XY model) are tabulated in Table \ref{table1}. 
The $1/q^2$-dependence of transition temperatures and exponents 
are shown in Fig.~\ref{fig_5}.  There, the exact results for 
$q=4$ are also given; that is, the Ising singularity at 
$T_c = [\ln(\sqrt{2}+1)]^{-1} = 1.1346$ with $\eta=1/4$. 
The transition temperature $T_1$ becomes smoothly lower with larger $q$; 
in the lowest order we find that $T_1 \propto 1/q^2$, 
which is consistent with the theoretical prediction \cite{Jose}. 
The critical exponent $\eta$ at $T_2$ is a universal constant, 
and compatible with the theoretical prediction $\eta=1/4$. 
The estimates of the critical exponent $\eta$ at $T_1$ remarkably 
coincide with the theoretical prediction $\eta=4/q^2$; 
1/9=0.111 for $q$=6, 1/16=0.0625 for $q$=8, and 
1/36=0.0278 for $q$=12. 
This is the {\it first} systematic report of confirming 
the theoretical prediction. 

\section{Summary and discussions}

To summarize, we have extended the PCC algorithm \cite{PCC} 
to the study of the XY and clock models. 
Wolff's idea of the embedded cluster formalism \cite{Wolff89} 
is used for treating the system with the vector order parameter. 
The KT transitions of the 2D XY and clock models are 
studied by using the FSS analysis 
based on the KT form of the correlation length. 
For dealing with the low-temperature transition temperature, 
$T_1$, we have employed a slightly modified algorithm. 
Investigating the $q=6,8,12$ clock models, we have 
systematically confirmed the prediction of $\eta(T_1)=4/q^2$. 
We have shown that small logarithmic corrections are present 
in the KT transitions.  The sign of the logarithmic-correction 
exponent $r$ is positive for all cases of the XY model and 
the clock models at both $T_1$ and $T_2$, which is compatible 
with Janke's result \cite{Janke97}, but different from 
the theoretical prediction $r=-1/16$ \cite{Kosterlitz}. 
The present precise numerical results may stimulate 
the refined renormalization-group study of the KT transitions. 

In the previous numerical studies of the KT transitions, 
one might resort to a big scale calculation using an extensive 
computer resource \cite{Gupta}, or one might use some special boundary 
conditions \cite{Olsson}.  It is due to the subtlety of 
the KT phase transitions; that is, $T_{\rm KT}$ is not the crossing point 
but the spray-out point of the existence probability or the Binder parameter. 
Moreover, the low-temperature transition $T_1$ for the clock model 
is difficult to study because the system is nearly ordered; 
it is more difficult for larger $q$. 
We should stress that using the present efficient method of 
numerical simulation, we have successfully made a systematic study 
of the XY and clock models with much less efforts.

Our formalism of the vector order parameter is easily extended to 
the general $O(n)$ model, where the percolation of $n$ types 
of Ising spins will be considered.  
Then, more problems of interest can be studied by the PCC algorithm. 
The PCC algorithm can be also applied to the quantum Monte Carlo simulation 
with the cluster algorithm \cite{Evertz93,Kawashima94}.  
It will be interesting to compare the present result with the 
quantum XY model.

\section*{Acknowledgments}

We thank N. Kawashima, H. Otsuka, M. Itakura, and Y. Ozeki 
for valuable discussions.  
Thanks are also due to M. Creutz for bringing our attention 
to the clock model having two phase transitions. 
The computation in this work has been done using the facilities of 
the Supercomputer Center, Institute for Solid State Physics, 
University of Tokyo. 
This work was supported by a Grant-in-Aid for Scientific Research 
from the Japan Society for the Promotion of Science.

\begin{table}
\caption{Transition temperatures and exponents $\eta$ 
for the 2D $q$-state clock model.}
\label{table1}
\begin{tabular}{lllll}
          & $T_2$   & $\eta(T_2)$ & $T_1$ & $\eta(T_1)$  \\
\tableline
$q=6$     & 0.9008(6) & 0.243(4) & 0.7014(11) & 0.113(3)  \\
$q=8$     & 0.8936(7) & 0.243(4) & 0.4259(4)  & 0.0657(2) \\
$q=12$    & 0.8937(7) & 0.246(5) & 0.1978(5)  & 0.0270(5) \\
{\rm XY} ($q=\infty$) & 0.8933(6)  & 0.243(4) & --------- & --------- \\
\end{tabular}
\end{table}

\end{multicols}

\end{document}